\documentclass[reprint,prb]{revtex4-2}
\usepackage[T1]{fontenc}
\usepackage{graphicx}
\usepackage{rotating}
\usepackage{bm}        
\usepackage{amssymb}   
\usepackage{bm}
\usepackage{float}
\usepackage{tikz}
\usepackage{diagbox}
\usepackage{braket}
\usepackage{color} 
\usepackage{silence}
\usepackage{colortbl}
\usepackage{amsmath}
\usepackage[colorlinks=true,linkcolor=blue,citecolor=blue]{hyperref}
\usepackage{amsmath,amssymb,amsthm, mathtools, enumitem, empheq, gensymb, textcomp, bm, booktabs, multirow}



\begin{document}

\preprint{APS/123-QED}

\title{Extrapolation of polaron properties to low phonon frequencies by Bayesian machine learning}
\author{Pranav Kairon}
\affiliation{Department of Physics and Astronomy, University of British Columbia, Vancouver, B.C., Canada}
\affiliation{Stewart Blusson Quantum Matter Institute, Vancouver, B.C., Canada.}%
\author{John Sous}
\affiliation{Department of Chemistry and Biochemistry, University of California, San Diego, La Jolla, California 92093, USA}
\author{Mona Berciu}
\affiliation{Department of Physics and Astronomy, University of British Columbia, Vancouver, B.C., Canada}%
\affiliation{Stewart Blusson Quantum Matter Institute, Vancouver, B.C., Canada}%
\author{Roman V. Krems}
\email{rkrems@chem.ubc.ca}
\affiliation{Department of Chemistry, University of British Columbia, Vancouver, B.C., Canada}%
\affiliation{Stewart Blusson Quantum Matter Institute, Vancouver, B.C., Canada}%



\date{\today} 

\begin{abstract}
Feasibility of accurate quantum calculations is often restricted by the dimensionality of the truncated Hilbert space required for the numerical computations. 
The present work demonstrates Bayesian machine learning (ML) models that use quantum properties in an effectively lower-dimensional Hilbert space to make predictions for the Hamiltonian parameters that require a larger basis set as applied to a classical problem in quantum statistical mechanics,  the polaron problem.
We consider two polaron models: the Su-Schrieffer-Heeger (SSH) model and the mixed SSH - Holstein model. We demonstrate ML models that can extrapolate polaron properties
in the phonon frequency. We consider the sharp transition in the ground-state momentum of the SSH polaron and examine the evolution of this transition from the anti-adiabatic regime to the adiabatic regime. 
We also demonstrate Bayesian models that use the posterior distributions of highly approximate quantum calculations as the prior distribution for models of more accurate quantum results. 
This drastically reduces the number of fully converged quantum calculations required to map out the polaron dispersion relations for the full range of Hamiltonian parameters of interest.
\end{abstract}

\maketitle


\section{Introduction}
The electron-phonon interactions in lattice systems give rise to polarons. These interactions can be represented by models that describe the effect of phonons on the on-site energy of the bare particle and models that modulate the amplitude for transitions between lattice sites. The notable examples are the Holstein model and the  Su-Schrieffer-Heeger model. 
The Holstein electron - phonon coupling originates from the change of the potential energy of an electron in a given lattice site due to lattice distortions  \cite{holstein1959studies1,holstein1959studies2}. 
The SSH coupling originates from the modulation of lattice site separations due to lattice vibrations  \cite{ssh_1,ssh_2,ssh_3}. Polarons described by these two models exhibit very different properties. 
For example, the effective mass of the Holstein polaron increases monotonically with the electron - phonon coupling strength, whereas the effective mass of the SSH polaron exhibits a sharp transition and can be low at strong coupling  \cite{DominicMona}.  It has been argued that the low effective mass of SSH polarons, and bipolarons \cite{PhysRevLett.121.247001}, may allow for high-temperature bi-polaronic superconductivity  \cite{polaron_1,polaron_2}.  Various extensions of these arguments have been considered to explore the properties of SSH-like polarons and bipolarons.  For example, Sous et al. \cite{PhysRevLett.121.247001} showed that a breathing-mode Peierls model leads to polarons whose effective mass remains approximately constant through transition to the strong coupling regime for a one-dimensional model. The same authors introduced the bond-Peierls model \cite{Bond_Peierls_polaron,polaron2D} to study this phenomenon in two dimensions, leading to the discovery of light polarons at strong coupling in two dimensions. This result was later extended to bipolarons in the presence of screened \cite{sous2022bipolaronic_prx} and unscreened Coloumb repulsion \cite{sous2022bipolaronic}, illustrating that phonons stimulate strongly bound yet light bipolarons, resilient to local repulsive interactions of the Hubbard type. 

However, these results, though encouraging for the prospects of bipolaronic high-temperature superconductivity \cite{sous2022bipolaronic_prx}, are based on numerical calculations in the anti-adiabatic limit, i.e. at high phonon frequencies, where phonon dynamics are fast compared to electron dynamics. The electron - phonon interactions for most materials are dominated by those in the adiabatic limit, i.e. the regime of low phonon frequencies. This limit is difficult to explore by numerical computations due to the explosion of the phonon Hilbert space \cite{QMC_for_eph,DMC_for_eph}. 
In practice, most numerical methods cannot probe the extreme adiabatic limit below $\Omega/t \lesssim 0.1$, where $\Omega$ is the phonon frequency and $t$ is the bare particle lattice hopping amplitude. 
Where possible, most numerical calculations for such phonon frequencies are restricted to
specific models with favorable local structure, such as the Holstein model \cite{ggce_main}. Ref. \cite{ggce_main} is an example of a fully converged quantum calculation
that uses  generalized Green's function cluster expansion (GGCE) to compute the properties of 
 the Holstein model at  $\Omega/t \lesssim 0.1$. 
Because lattice vibrations are extremely slow in this extreme adiabatic limit, one may treat phonons classically \cite{kim2023semi}.  
However, in many materials,  multiple phonon branches with varying energies couple to electrons making the classical treatment cumbersome. 
Therefore, it is important to understand quantum corrections 
and effects of non-local electron - phonon couplings in the adiabatic limit, even for simplified models with a single phonon branch that serve as a reference for more complex models. 
It is not understood if the sharp transition observed in the effective mass for the SSH polaron \cite{DominicMona} survives
and whether the SSH bipolarons with low effective mass \cite{PhysRevLett.121.247001} may form at strong coupling in the adiabatic regime.


Motivated by this uncertainty and the difficulty of numerical analysis of polarons at low phonon frequencies, the present work explores the possibility of building machine learning (ML) models that can be trained by either (i) the polaron properties at high phonon frequencies or (ii) low-level, highly approximate, quantum calculations, in order to make accurate predictions of polaron properties at low phonon frequencies. 
In recent years, ML has become a powerful tool for numerical predictions for many applications in condensed matter physics. ML algorithms have been used for accelerating DFT calculations \cite{PhysRevMaterials.6.040301}, exploring phase diagrams \cite{carrasquilla2017machine,van2017learning, PhysRevX.7.031038, PhysRevB.96.184410}, determining the wavefunction in variational approaches with sign problems and sampling configurations from many-body Hamiltonians \cite{PhysRevX.10.041026, PhysRevResearch.3.043126, PhysRevB.95.041101}. 
Bayesian kernel-based ML models are particularly well-suited for data-starved models \cite{NEURIPS2020_a9df2255} and where predictions are required outside the range of training data  \cite{rodrigo_roman,Dai_2023,Dai_krems_qk,PhysRevResearch.2.032051,Asnaashari_2022,jie_krems,jie_prl,Vargas_Hernandez2020}.
For example, our previous work demonstrated that Bayesian ML models  can be used to identify multiple phase transitions using system properties only from one phase \cite{rodrigo_roman}. 

Here, we extend the work in Ref. \cite{rodrigo_roman} to problems with more dimensions with a particular focus on extrapolation in the phonon frequency domain. 
Our ML models are based on Gaussian Process regression (GPR) trained by the polaron properties computed using the GGCE method \cite{ggce_main, ggce_code}. 
Our goal is to develop machine learning (ML) models that use polaron properties in the anti-adiabatic limit to predict polaron properties in the adiabatic limit. 
The resulting ML models are shown to capture the evolution of the SSH polaron properties down to phonon frequencies, where polaron dispersion calculations are currently out of reach of rigorous quantum theory. 
We identify the evolution of the position of the sharp change in the SSH polaron effective mass in the Hamiltonian parameter space with the change of the phonon frequency and the accuracy level of the GGCE calculations.


The paper is organized as follows: Section \ref{sec-model_ggce} outlines the electron-phonon model and the GGCE method used to calculate phonon dispersions. Section \ref{sec-gpml} describes the algorithms used for the ML predictions. 
Section \ref{sec-gpml-2} discusses how to construct multi-fidelity models for improving the accuracy of quantum predictions. In Section \ref{sec-results}, we present the main findings and results obtained for different electron-phonon models. We discuss the specific implications of our work for 
the ongoing debate about the survival of the sharp transition of the SSH polaron effective mass in the extreme adiabatic regime. 
Section \ref{sec-conclusion} concludes by a  summary of key results and implications.

\section{Green's function calculations}

\label{sec-model_ggce}

The most general Hamiltonian we consider in this work can be written as follows: 
\begin{equation}{\label{hamiltonian}}
\begin{aligned}
\hat{H}= & -t \sum_{\langle i j\rangle} \hat{c}_i^{\dagger}
\hat{c}_j + 
\Omega \sum_i  \hat{b}_i^{\dagger} \hat{b}_i +
 \alpha_{\mathrm{H}} \sum_i \hat{c}_i^{\dagger} 
 \hat{c}_i\left(\hat{b}_i^{\dagger}+\hat{b}_i\right)
\\
& + \alpha_{\mathrm{SSH}} \sum_{\langle i
j\rangle}\left(\hat{c}_i^{\dagger} \hat{c}_j+\hat{c}_j^{\dagger}
\hat{c}_i\right)\left(\hat{b}_i^{\dagger}+\hat{b}_i-\hat{b}_j^{\dagger}-\hat{b}_j\right) 
\end{aligned}
\end{equation}
where $t$ is the electron hopping amplitude,
$\Omega$ is the dispersionless frequency of phonons, $\alpha_{\rm SSH}$ and $\alpha_{\rm H}$ are the corresponding strengths of the electron-phonon couplings for the SSH and Holstein models, respectively, and the operators $\hat c_i$ and $\hat b_i$ create the electron and phonons at the site $i$ of an infinite lattice. 
In the present work, we consider a single phonon branch and perform calculations for two types of models: (i) the pure SSH model that corresponds to the  $\alpha_{\rm H} = 0$ limit of Eq. (\ref{hamiltonian}); and (ii) the mixed model (\ref{hamiltonian}) with both $\alpha_{\rm SSH} \neq$ and $\alpha_{\rm H} \neq 0$.

The electron - phonon coupling is quantified by $\lambda_{\rm SSH}=2\alpha_{\rm SSH}^{2}/\Omega t$ for the pure SSH model and by a combination of $\lambda_{\rm SSH}$ and $\lambda_{\rm H}=\alpha_{\rm H}^{2}/2\Omega t$ for the mixed model. 
We also use the adiabaticity ratio 
\begin{equation}
\Lambda=\Omega/4t,
\end{equation}
 to quantify the regime of electron - phonon interactions.

We compute the Green's function $G(k,\omega)$ of the phonon-dressed particle 
with the GGCE method \cite{ggce_code,ggce_main}, which uses the momentum average (MA) approximation \cite{mona_ma1,mona_ma2,mona_ma3,mona_ma3_ssh}  along with a variational ansatz 
parameterized by the spatial extent of the phonon cloud ($M$) and the number of bosons in a single phonon cloud ($N$). The convergence in the limit of $M\rightarrow \infty, N \rightarrow \infty$
corresponds to the exact solution.
All computations in the present work are for $T=0$. 
The Green's function is computed on a grid of $\omega$ and polaron momentum $k$. The numerical complexity of the calculations scales combinatorially with $M$ and $N$.

\begin{figure*}
            \includegraphics[width=0.98\linewidth]{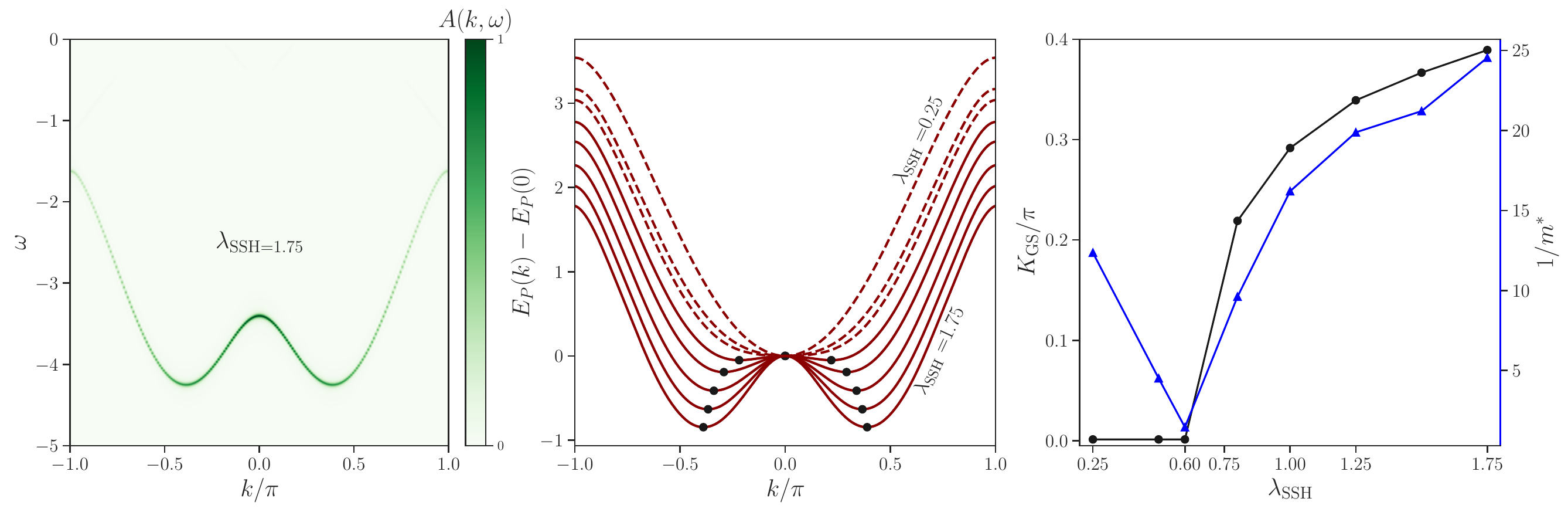}
        \caption{(Left) Spectral function $A(k,\omega)$ for $\Omega=4.0$ and
        coupling strength $\lambda_{\rm SSH}=1.75$ and $\alpha_{\rm H}=0$; (Center) Polaron energy $E_{P}(k)$
        for several coupling strengths: dotted curves are polaron energies for
        $\lambda < \lambda^{\rm c}$ with the energy minimum at $k=0$ (black
        circles) and solid curves are for $\lambda < \lambda^{\rm c}$ with ground
        state occurring at $k=\pm K_{\rm GS}$ (black cirlces).  (Right) SSH polaron ground
        state momentum $K_{\rm GS}$ and effective polaron $1/m^*$ as functions of
        $\lambda_{\rm SSH}$. The sharp transition occurs at
        $\lambda^{\rm c}_{\rm SSH}=0.644$. \\
        }
        \label{fig_theory}
\end{figure*}

To obtain the polaron dispersions, we calculate the momentum-frequency resolved spectral function $A(k,\omega)$ defined as $A(k,\omega) \equiv -\frac{1}{\pi}\text{Im } G(k,\omega)$. We follow the evolution of the lowest peak in $A(k,\omega)$ which corresponds to the polaron band for all $k$ values, and obtain the polaron dispersion for a range of coupling strengths $\lambda_{\rm SSH}$. For a given phonon frequency $\Omega$, we compute the polaron ground state momentum $K_{\rm GS}$ as a function of $\lambda_{\rm SSH}$ and observe the sharp transition $K_{\rm GS} = 0 \leftrightarrow K_{\rm GS} > 0$ at a critical coupling strength $\lambda_{\rm SSH}^{c}$  \cite{DominicMona}.   Fig. \ref{fig_theory} illustrates the spectral function $A(k, \omega)$, the energy disperion and the evolution of $K_{\rm GS}$ thus computed for the SSH polaron. In particular, it demonstrates the sharp transition in $K_{\rm GS}$, which also manifests itself as a sharp transition in the effective mass of the polaron
\begin{equation}
    {1/m^*} = \biggr[\frac{d^2 E_P(k)}{dk^2} \biggr]_{K_{\rm GS}},
\end{equation}
also shown in the right panel of Fig. \ref{fig_theory}.

\section{Bayesian model construction}\label{sec-gpml}

We use Gaussian process regression (GPR) to build machine learning models in the present work. 
GPR is a supervised learning algorithm trained to predict a continuous variable by a normal distribution of functions ${f}(\bm x_*)$. 
The training data set
comprises $\mathcal{D} = \{ \mathbf{X}, {\bm y} \}_{i=1}^{n}$, with $n$ $p$-dimensional input vectors denoted as $\mathbf{X} = [{\bm x}_1, {\bm x}_2, \ldots, {\bm x}_n]^{\intercal}$ and the corresponding values of the  output variable collected in the vector $\bm y$. 
The mean of the predictive distribution is given by
\begin{equation}\label{gp_eq}
{f}(\bm x_*) = {\bm k}_{*}^{T} (\mathbf{K}+\sigma^{2} {\bf I})^{-1}{\bm y}
\end{equation}
where $\mathbf{K}$ is a symmetric positive definite kernel matrix of size $n \times n$, computed over the training set,  ${\bm k}_{*}^{T}$ is a vector of length $n$, with elements representing the kernel function evaluated for the combinations of ${\bm x_i} \in {\bf X}$ and arbitrary ${\bm x}_*$,
 $\bf{I}$ is the identity matrix and $\sigma^{2}$ is a hyperparameter that accounts for noise in the dataset. In the present work, we set $\sigma^{2}$ to zero. 
The elements of $\bf K$ and ${\bm k}_{*}^{T}$ are given by the values of the kernel function ${\kappa }({\bm x},{\bm x}')$. 
A GP model is trained by maximizing the logarithm of the marginal likelihood (LML) 
\begin{equation}
    \text{log}~\mathcal{L}(\mathbf{\bm \theta})= -\frac{1}{2}{\bm y}^{T}\mathbf{K}^{-1}{\bm y} -\frac{1}{2}\text{log}|\mathbf{K}|-\frac{n}{2}\text{log}~2\pi
\end{equation}
by varying the parameters of the kernel function represented collectively by $\bm \theta$.

The functional form of  the kernel function defines the model $\mathcal{M}_i$.
The choice of the kernel function is critically important for data-starved problems and for extrapolation problems.
To identify the optimal functional form of the kernel function, we follow the approach developed by Duvenaud et al. \cite{ML_bic_1, ML_bic_2}.

The specific implementation of the kernel selection algorithm is described in detail elsewhere \cite{ML_bic,ML_bic_1,ML_bic_2}.
In brief, the algorithm begins with the following set of kernel functions (base kernels): 
    \begin{align}
        \label{base1}
        & \kappa\left({\bm x}, {\bm x}^{\prime}\right)=\sigma \exp \left(-\frac{1}{2} r^2\left({\bm x}, {\bm x}^{\prime}\right)\right)\\
        & \kappa({\bm x}, {\bm x}') = \sigma \left(1 + \sqrt{5} \cdot r({\bm x}, {\bm x}') + \frac{5}{3} \cdot r^2({\bm x}, {\bm x}')\right) \nonumber \\ 
        \label{base2}
        & \quad \times \exp\left(-\sqrt{5} \cdot r({\bm x}, {\bm x}')\right) \\ 
        & \kappa\left({\bm x}, {\bm x}^{\prime}\right)=\sigma\left(1+\frac{1}{2 \alpha} r^2\left({\bm x}, {\bm x}^{\prime}\right)\right)^{-\alpha}
\label{basekernels}
    \end{align}
where $r^2\left({\bm x}, {\bm x}^{\prime}\right)=\left({\bm x}-{\bm x}^{\prime}\right)^T \times {\bf \Lambda} \times\left({\bm x}-{\bm x}^{\prime}\right)$, $\bf \Lambda$ is a diagonal matrix of shape $|\mathcal{M}_i| \times |\mathcal{M}_i|$,
and $|\mathcal{M}_i|$ is the length of the kernel parameter vector for the model $\mathcal{M}_i$. 

To minimize the number of free parameters in the kernel function,  one often uses isotropic kernels, such that ${\bf \Lambda} = l \times \bf{I}$, where $l$ is a scalar. 
 In this work we find that the kernel anisotropy is important for the performance of the resulting models. We therefore
  use anisotropic kernels, with different parameters for different input features.
  We start with the simple kernels given by Eqs. (\ref{base1}) -- (\ref{basekernels}).
These base kernels are combined into products and linear combinations. The optimal combination of the base kernels is then combined with each of the base kernels and the process is iterated.   
At each step of the kernel selection, the model ${\cal M}_i$ is optimized by maximizing LML and 
the optimal model is selected by the value of the Bayesian information criterion \cite{ML_bic}:
\begin{equation}
    \text{BIC}(\mathcal{M}_i)=\text{log}~\mathcal{L}(\mathbf{{ \hat {\bm \theta}}}_i)-\frac{1}{2}|\mathcal{M}_i|~\text{log}~n
\end{equation}
where $\hat {\bm \theta}$ is a vector collecting the optimal values of the parameters of the kernel function ${\cal M}_i$. 
We refer to GP kernels thus constructed as composite kernels. For most of the calculations presented in the subsequent section, the most optimal kernel function was identified to be the product of Eq. (\ref{base1}) and Eq. (\ref{base2}).

\begin{figure*}
        \includegraphics[width=0.45\textwidth]{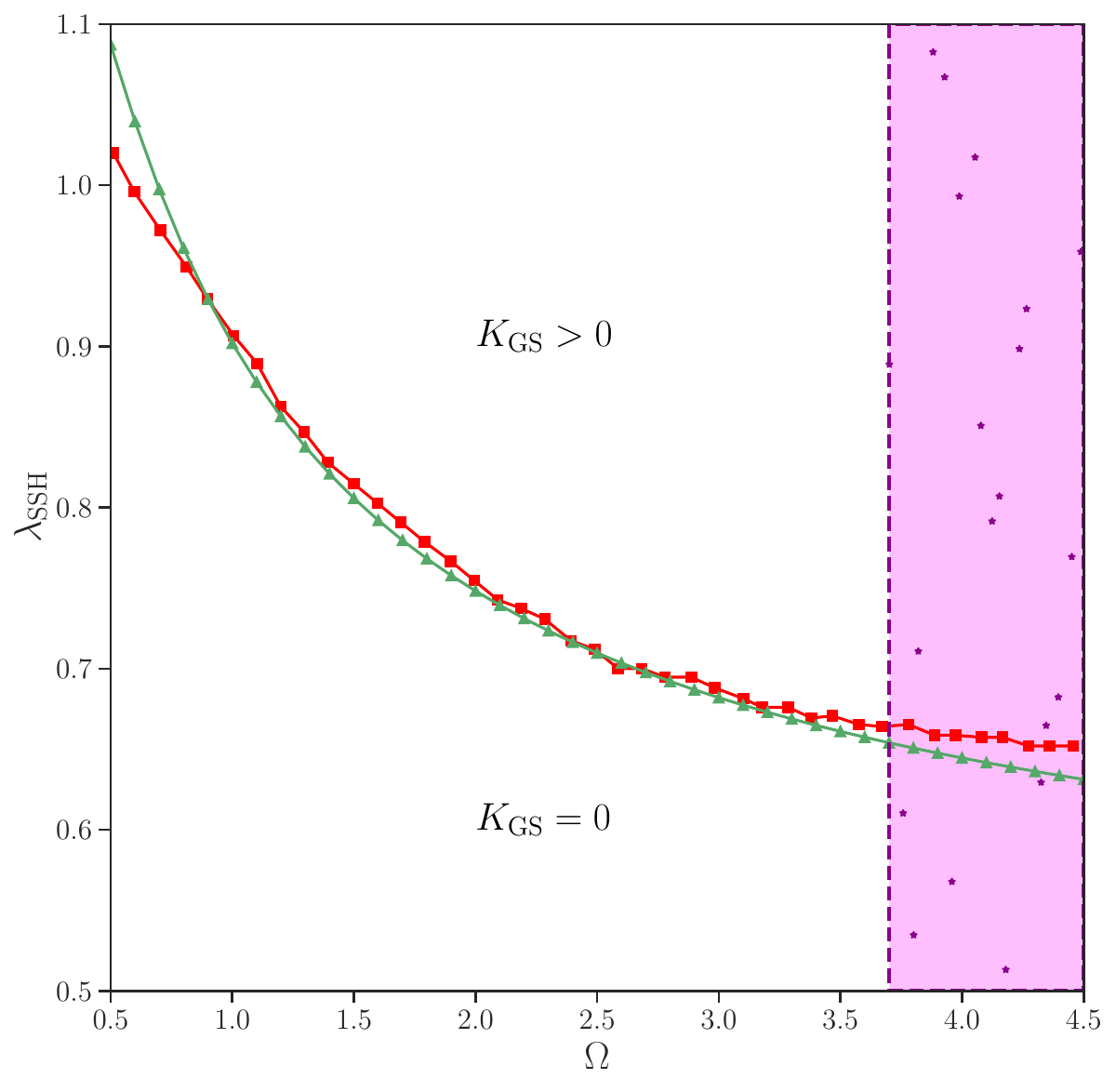}
        \includegraphics[width=0.45\textwidth]{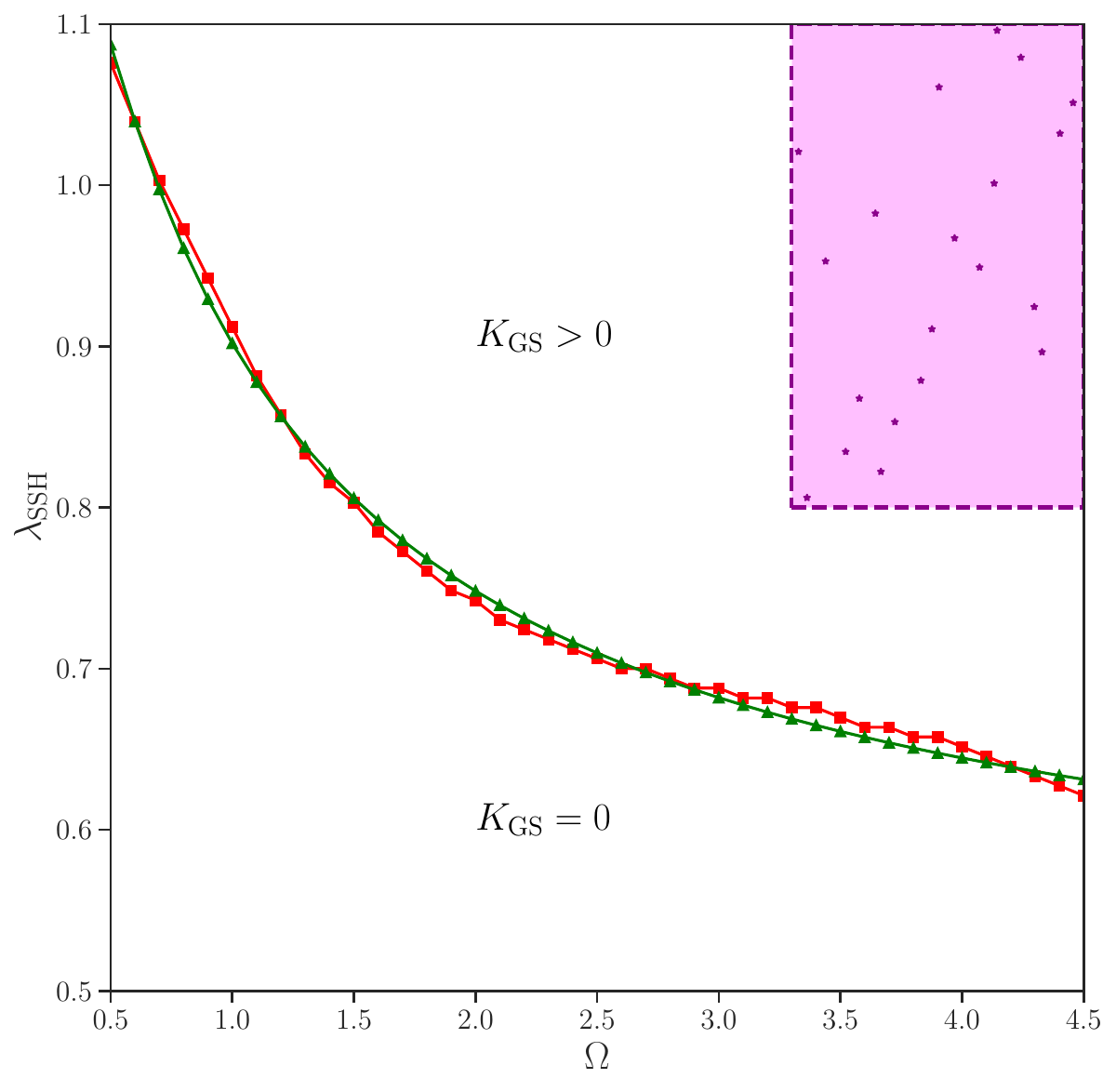}
        \label{fig:ssh_2}  
  \caption{Ground state momentum $K_{\rm GS}$ of the SSH polaron as a function of $\lambda_{\rm SSH}$
  and phonon frequency ($\Omega$): squares -- ML predictions; triangles -- GGCE calculations. 
  The ML models are trained by 600 polaron energies randomly sampled from the three-dimensional space $[\Omega, \lambda_{\rm SSH}, k]$ indicated by the symbols in the shaded region. 
    }
  \label{fig:ssh_main}
\end{figure*}

\subsection{Multi-Fidelity learning}{\label{sec-gpml-2}}

As mentioned above, the complexity of Green's function calculations increases rapidly with $N$ and $M$. One of the goals of the present work is to make accurate predictions of polaron properties using highly approximate, but inexpensive, calculations with small $N$ and $M$. To this end, we extend the algorithm described in the previous section to produce models trained by a combination of a large set of inexpensive results (low $N$ and $M$) and a small number of converged calculations (large $N$ and $M$). The assumption is that approximate calculations with small $N$ and $M$
are correlated with fully converged results. The goal is to learn these correlations for extrapolation models. The possibility of such models in the context of Bayesian machine learning was previously demonstrated, for example,  by Jie and Krems \cite{jie_prl} and Jasinski et al. \cite{PhysRevResearch.2.032051}. For example, Ref.  \cite{jie_prl}  demonstrated Bayesian models that can learn from a large  number of classical dynamics calculations and a small number of quantum dynamics calculations in order to make accurate quantum predictions, including predictions of quantum resonances that are completely absent in classical results.  Similar approaches have also been used for improving the accuracy of  potential energy surfaces for chemical dynamics applications \cite{mult_fid_chem}.

In the present work, we follow the approach described by Perdikaris et al. \cite{perdikaris2017nonlinear} based on non-linear auto-regressive multi-fidelity Gaussian processes (NARGP). The goal is to build a model of a target function $f_{t}$, which is expensive to evaluate, by exploiting cheap   auxiliary functions $f_0,f_1...,f_{t-1}$ which approximate $f_{t}$ in the increasing order of fidelity. The relationship between the target function and the auxiliary functions can be generally represented as
\begin{equation}
    f_{\text{t}}(\bm x) = \rho_{t-1}( \, f_{\text{t-1}}(\bm x)) + \delta_{t}(\bm x) 
\end{equation}
where $\rho_{t-1}(\cdot)$ is a nonlinear mapping between two successive auxiliary functions, which  is independent of $\delta_{t}(x)$.

The covariance structure of a GP that models this family of functions can be written as:
\begin{eqnarray}
    k_{t}({\bm x},{\bm x}')=k_{z_{t-1}}({\bm x},{\bm x}') \times k_{f_{t-1}}(f^{*}_{t-1}({\bm x}),f^{*}_{t-1}({\bm x}'))\nonumber \\ +k_{\gamma_{t}}({\bm x},{\bm x}')~~~~~~
\end{eqnarray}
When training the GP model by LML maximization, the GP prior of $f_{t-1}$ is replaced with the GP posterior $f^*_{t-1}$ obtained using the previous fidelity level. 
The expectation that the resulting Bayesian models must be efficient is based on the assumption that low-level approximate calculations provide better description of the physical process than random guessing. 

NARGP models can also be written as:
\begin{equation}
    f_{t}({\bm x})=g_{t}({\bm x},f^*_{t-1}({\bm x}))
\end{equation}
where $g_t \sim \mathcal{GP}(f_t|0,k_{t}(({\bm x},f^{*}_{t-1}({\bm x})),({\bm x}',f^{*}_{t-1}({\bm x}'),{\bm x})))$. Using NARGP, one can propagate the uncertainty fully from a lower level of fidelity to  higher levels of fidelity. Hence, training a NARGP model is akin to training a regular GP model. For a more detailed discussion on multi-fidelity Gaussian processes see Ref. \cite{multifidelityreview}.

\section{\label{sec-results}Results}
We present the main results of this work as follows. First, we consider a pure SSH model and demonstrate ML models that extrapolate the ground state momentum of the SSH polaron in phonon frequency. We build several ML models trained by the polaron energy dispersions in different regimes of adiabaticity. We then apply a similar analysis to the combined  SSH - Holstein model. We obtain ML predictions of the sharp transition in the polaron ground-state momentum down to the extreme adiabatic limit. Finally, we demonstrate the application of multi-fidelity models to obtain accurate predictions of polaron properties from low-level GGCE calculations, thus reducing the overhead of computationally expensive numerical methods.

\subsection{SSH polaron}

We begin by training Gaussian process regression models with composite kernels by polaron energies $E_P(k, \Omega, \lambda_{\rm SSH})$ 
calculated using GGCE as described in Section \ref{sec-model_ggce}. The resulting ML models predict the polaron energy for given values of the Hamiltonian parameters $\Omega,\lambda_{\rm SSH}$ and a given value of $k$.
Our ML models can thus be viewed as surrogate models of GGCE calculations, with inputs given by three-dimensional vectors of $\left [ \Omega,\lambda_{\rm SSH},k\right]$ and outputs representing the polaron energy. 
In Section \ref{sec-results_d} below, we show how to extend these models to include the parameters $N$ and $M$ of the GGCE calculations on input, in order to improve the accuracy of predictions based on low-level GGCE calculations. 
In Ref. \cite{rodrigo_roman}, the ML models were trained by full polaron dispersions at fixed values of phonon frequency and particle - phonon couplings. We note that we use a different approach in the present work. 
In particular, we do not discriminate between the input variables $\left [ \Omega,\lambda_{\rm SSH},k\right]$. Instead, the polaron energies are computed at randomly selected combinations of the three variables $\left [ \Omega,\lambda_{\rm SSH},k\right]$. We use a Latin Hypercube sampling strategy \cite{loh1996latin} to avoid accidental clustering of the training data and train ML models by energy points in the three-dimensional space thus sampled.

Figure \ref{fig:ssh_main} illustrates the ML predictions of $K_{\rm GS}$ for the SSH polaron as a function of $\Omega = 0.5$ and $\lambda$ in the range up to $1.1$. 
Our particular focus is on the location of the sharp transition $K_{\rm GS} = 0 \leftrightarrow K_{\rm GS} >0$ in the $[\lambda_{\rm SSH}, \Omega]$ diagram. 
The models are trained by polaron energies computed for the Hamiltonian parameters in the anti-adiabatic limit $\Lambda>1$ indicated by the symbols in the shaded region.  The predictions are tested at different phonon frequencies, extending to the adiabatic regime. Figure \ref{fig:ssh_main} (left) illustrates ML extrapolation of the sharp transition in the frequency domain. It can be observed that ML models capture the evolution of the combination of $\lambda_{\rm SSH}$ and $\Omega$ that gives rise to the sharp transition for the SSH polaron down to low values of frequency $\Omega = 0.5$.

Figure \ref{fig:ssh_main} (right) demonstrates the accuracy of the ML predictions obtained by extrapolation in both $\Omega$ and $\lambda$. As illustrated by the shaded region of Figure \ref{fig:ssh_main} (right), the polaron energies used for training ML models are entirely in the high-frequency, $K_{\rm GS}>0$ region of the $[\lambda_{\rm SSH}, K_{\rm GS}]$ diagram, far removed from the location of the sharp transition. 
The evolution of the SSH polaron dispersion in the $[\lambda_{\rm SSH}, K_{\rm GS}]$ space is analogous to the evolution of free energy across second order phase transitions for some thermodynamic systems, such as the many-body spin system described by the Heisenberg model.  
Figure \ref{fig:ssh_main} (right) thus illustrates that ML models can be used to identify a phase transition by extrapolation of smooth functions from a single phase. 


\begin{figure}
  \includegraphics[width=0.9\columnwidth]{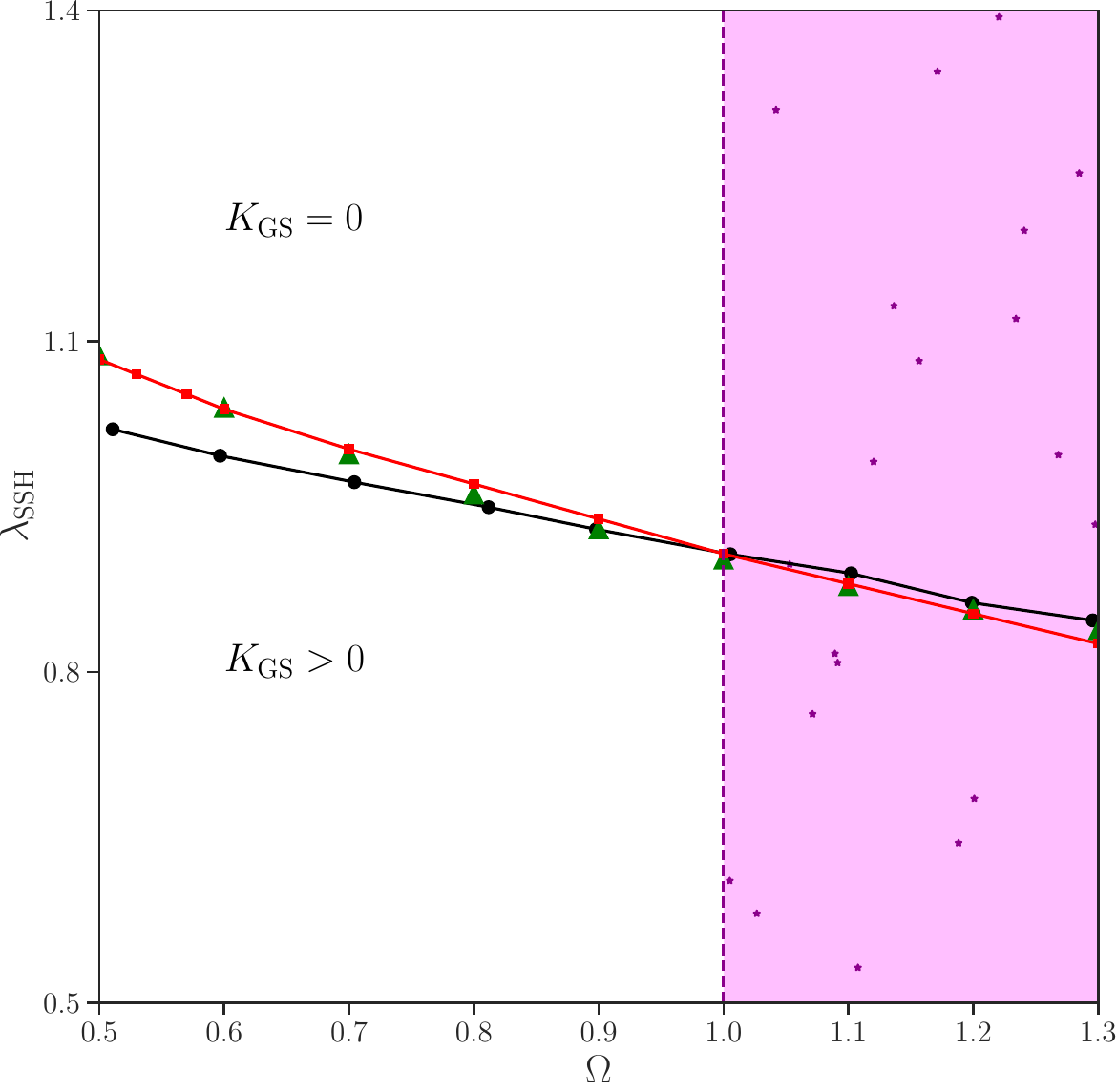}
  \caption{Ground state momentum $K_{\rm GS}$ of the SSH polaron as a function of $\lambda_{\rm SSH}$
  and phonon frequency ($\Omega$): squares -- predictions of ML models trained by polaron energies at phonon frequencies
  $\Omega \geq 1$; circles -- predictions of ML models trained by polaron energies at phonon frequencies
  $\Omega \geq 3.5$;
   triangles -- GGCE calculations. 
  The ML models are trained by 415 polaron energies randomly sampled from the three-dimensional space $[\Omega, \lambda_{\rm SSH}, k]$. The training range for predictions represented by squares is shown by the shaded region.
  }
\label{fig-ssh_2}
\end{figure}

The accuracy of ML predictions obtained by extrapolation from a given phase is expected to improve as the training data approach the transition. This is illustrated in Figure \ref{fig-ssh_2}, where the polaron energies used for training the ML models are computed in the phonon frequency range $\Omega \in [1, 1.3]$. This circles in this figure are the predictions of the ML model represented by squares in Figure  \ref{fig:ssh_main} (left).



\begin{figure}
  \includegraphics[width=0.95\columnwidth]{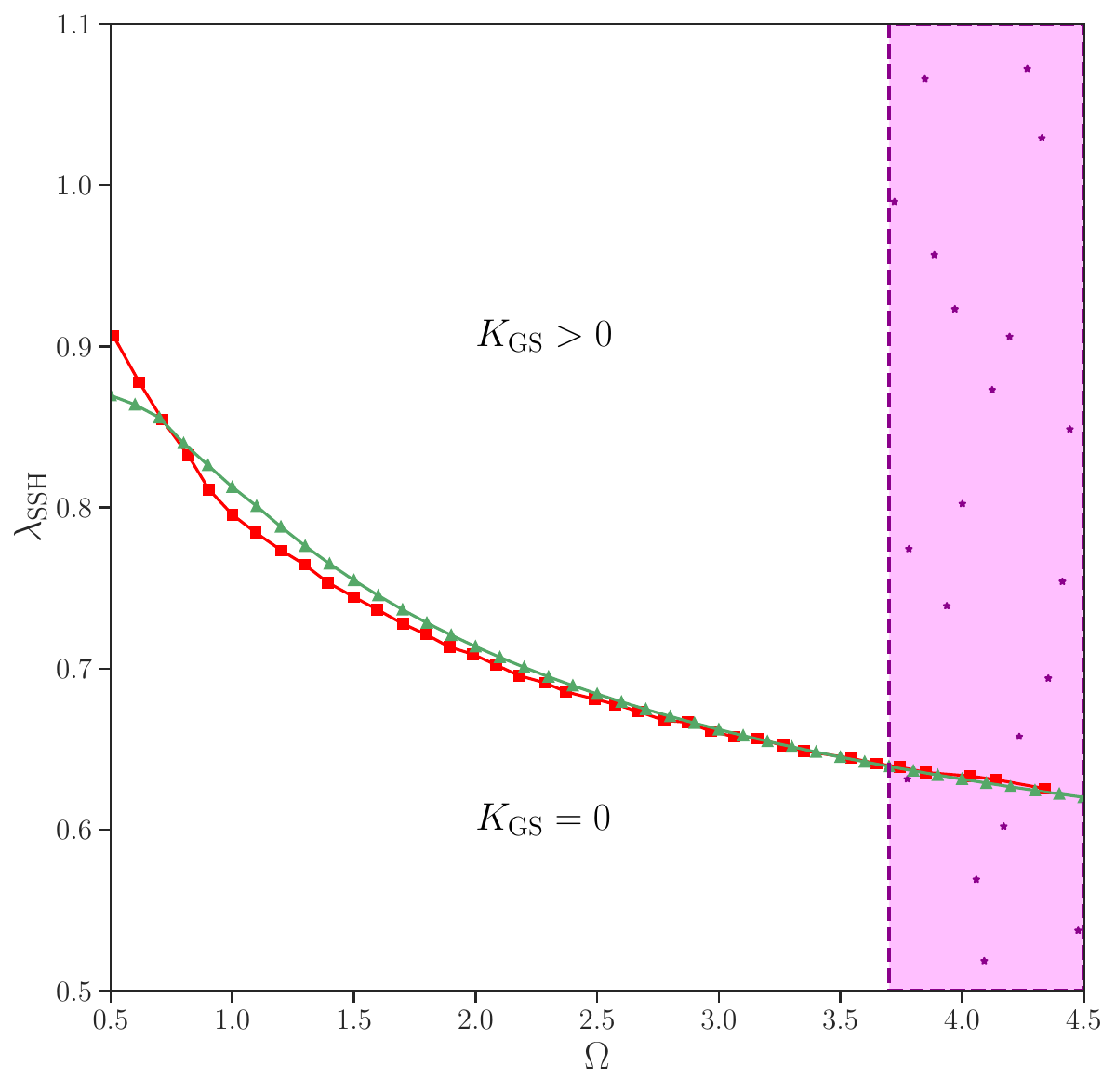}
  \caption{Ground state momentum $K_{\rm GS}$ of the polaron for the mixed Holstein - SSH model as a function of $\lambda_{\rm SSH}$
  and phonon frequency ($\Omega$): squares -- ML predictions; triangles -- GGCE calculations. 
The ML models are trained by 600 polaron energies randomly sampled from the three-dimensional space $[\Omega, \lambda_{\rm SSH}, k]$ indicated by the symbols in the shaded region.   The Holstein coupling is fixed to $\lambda_H=0.5$.
 }
\label{SSH-H}
\end{figure}

\subsection{Mixed Holstein -- SSH Model}
We now perform the same analysis as in the previous subsection but for a mixed coupling model given by Eq. (\ref{hamiltonian}). 
This model has been used for the description of excitations in molecular complexes and organic crystals \cite{ho_ssH_1,ho_ssh_2,ho_ssh_3}. It can be viewed as a generalization of both the SSH and Holstein models to include competition between phonon-induced interactions that modultae the potential and kinetic energy of the bare particle. 

It was previously shown that the ground-state momentum of the polaron described by the mixed model also exhibits a sharp transition. 
 However, the critical coupling strength for the generalized model is different. The dual coupling model gives rise to a critical coupling surface $\mathbf{\lambda^c}=[\lambda_{\rm SSH}^c, \lambda_H]$, where $\mathbf{\lambda^c}$ is a function of $\Omega$.
 We examine cuts through this surfaces that correspond to fixed Holstein coupling. 
  For the case illustrated in Fig. \ref{SSH-H}, we fix $\lambda_{\rm H}=0.5$. We train the ML models by data at $\Omega=[3.7,4.5]$ and predict the shift of this new critical coupling strength with the $\Omega$ phonon frequency. Here the polaron energies are calculated using MA \cite{mona_ma1}, via a single phonon branch.

\subsection{\label{sec-results_c} Predictions by extrapolation to extreme adiabatic regime}
The ML illustrated in the previous section are here used to predict the evolution of the sharp transition to the extreme adiabatic regime ($\Lambda>>1$). 
As $\Lambda$ increases, accurate calculations of the polaron energy become exceedingly difficult. 
As a result, $\Omega = 0.5$ is the lowest phonon frequency, for which polaron dispersions of the SSH polaron have been reported in the literature to date \cite{DominicMona}. Very recently, Grundner et al performed DMRG calculations for a lattice with 256 sites and the SSH particle - phonon coupling for the phonon frequency 0.2  \cite{grundner2023cooper}. Interestingly, the results of their work indicate the absence of the sharp transition at this phonon frequency.  
We attempt to explore the persistence of the sharp transition in the SSH polaron diagram by extrapolation with ML to $\Omega \leq 0.4$.

\begin{figure}
  \centering
  \includegraphics[width=0.95\columnwidth]{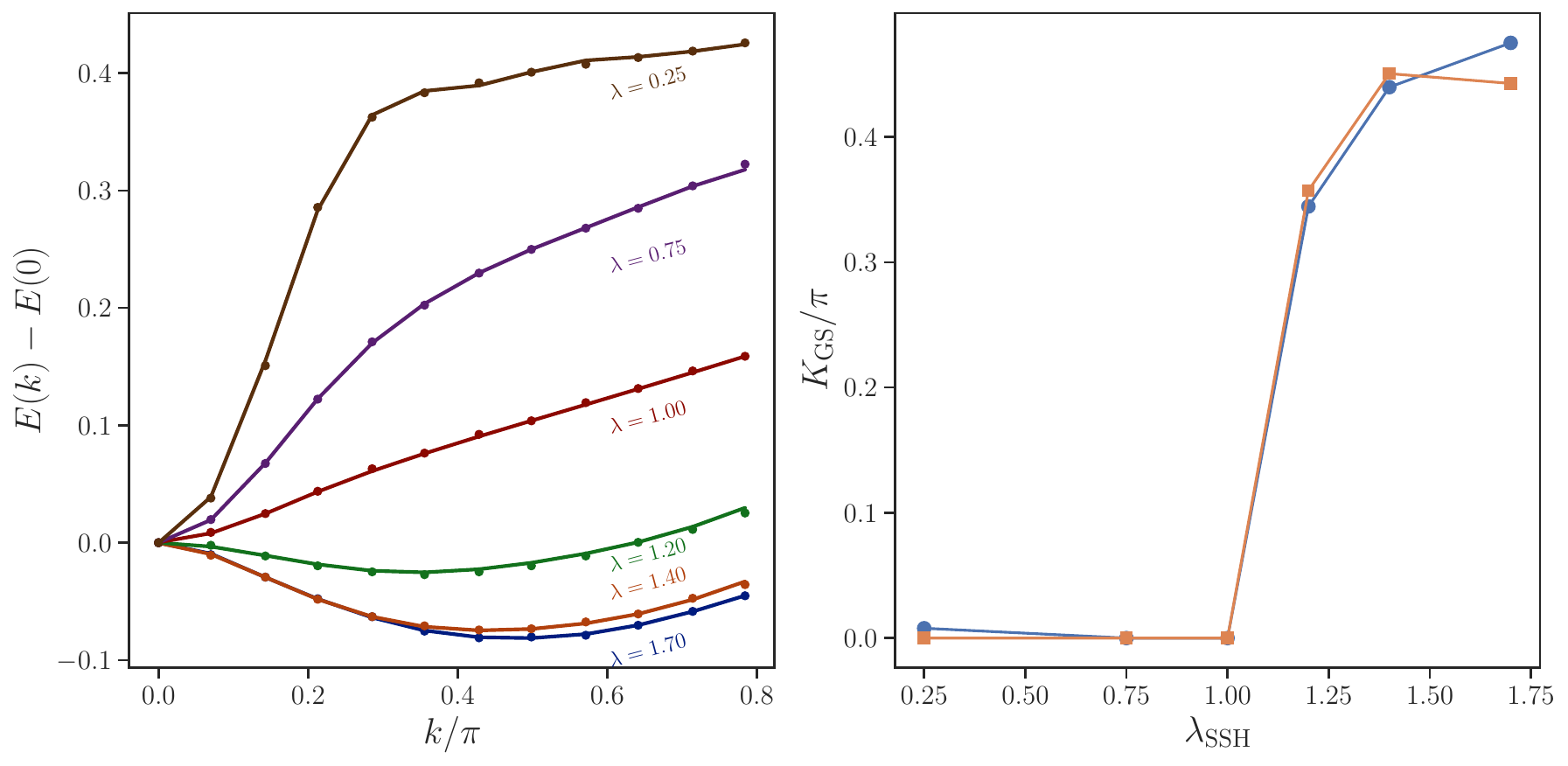}
  \caption{(Left) SSH polaron dispersions for $\Omega=0.4$: the symbols represent the results of the GGCE calculations with $M=5, N=10$; the solid lines are the ML predictions. The ML model is trained by the polaron dispersions at $\Omega \times \lambda_{\rm SSH}=[0.5,0.7] \times [0.25,1.7]$. 
  (Right) The sharp $K_{\rm GS} = 0 \leftrightarrow K_{\rm GS} >0$ transition observed in the ground state momentum at $\Omega=0.4$: circles -- the ML predictions; squares -- the GGCE calculations.}
  \label{ssh_low_04}
\end{figure}

First, we consider the case of $\Omega = 0.4$. 
Our models are trained by the polaron dispersions at $\Omega \times \lambda_{\rm SSH}=[0.5,0.7] \times [0.25,1.7]$, down to the lowest phonon frequencies probed by the previous calculations, and predict the polaron dispersions and transition curve at $\Omega=0.4$. 
The results reported in Fig. \ref{ssh_low_04}  show that ML models predict the expected trend for both the polaron dispersion and the evolution of the transition. The predicted value of $\lambda_{\rm SSH}^c$ is $1.19$. 

To verify this prediction, we perform selected GGCE calculations with $M=5, N=10$. The results are compared with the ML predictions in Fig. \ref{ssh_low_04}. 
The value of $\lambda_{\rm SSH}^c$ inferred from the GGCE calculations is between 1.1 and 1.2. A more accurate prediction would require a large number of GGCE calculations on a dense grid of 
$\omega,k,\lambda_{\rm SSH}$, which is currently out of reach of our computation resources.

\begin{figure*}
        \includegraphics[width=0.9\textwidth]{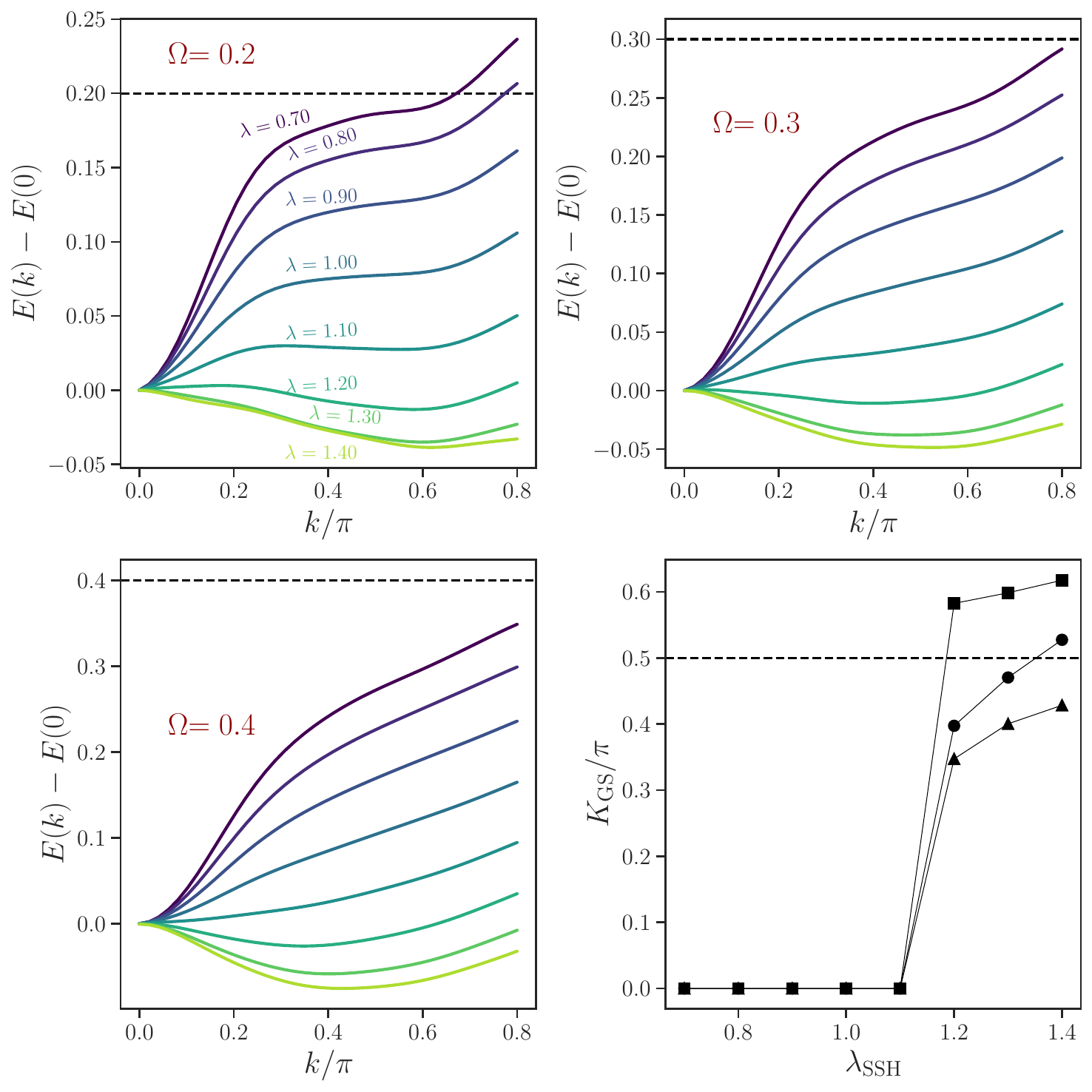}
\caption{Extrapolation of the polaron dispersion relation $E_{\rm P}$ in phonon frequency space ($\Omega$) for the pure SSH model. The dispersion curves are predicted by ML models trained at phonon frequencies $\Omega \in [0.4,0.9]$. The solid lines are the predictions of the model for various coupling strengths. The horizontal dashed lines show the limit of the dispersion bandwidth equal to $\Omega$. Lower right: The dependence of $K_{\rm GS}$ of the SSH polaron on $\lambda_{\rm SSH}$ at three phonon frequencies corresponding to the dispersion curves shown in the other panels.
}
  \label{extreme_adiabtic-1}
\end{figure*}

\begin{figure*}
       \includegraphics[width=0.9\textwidth]{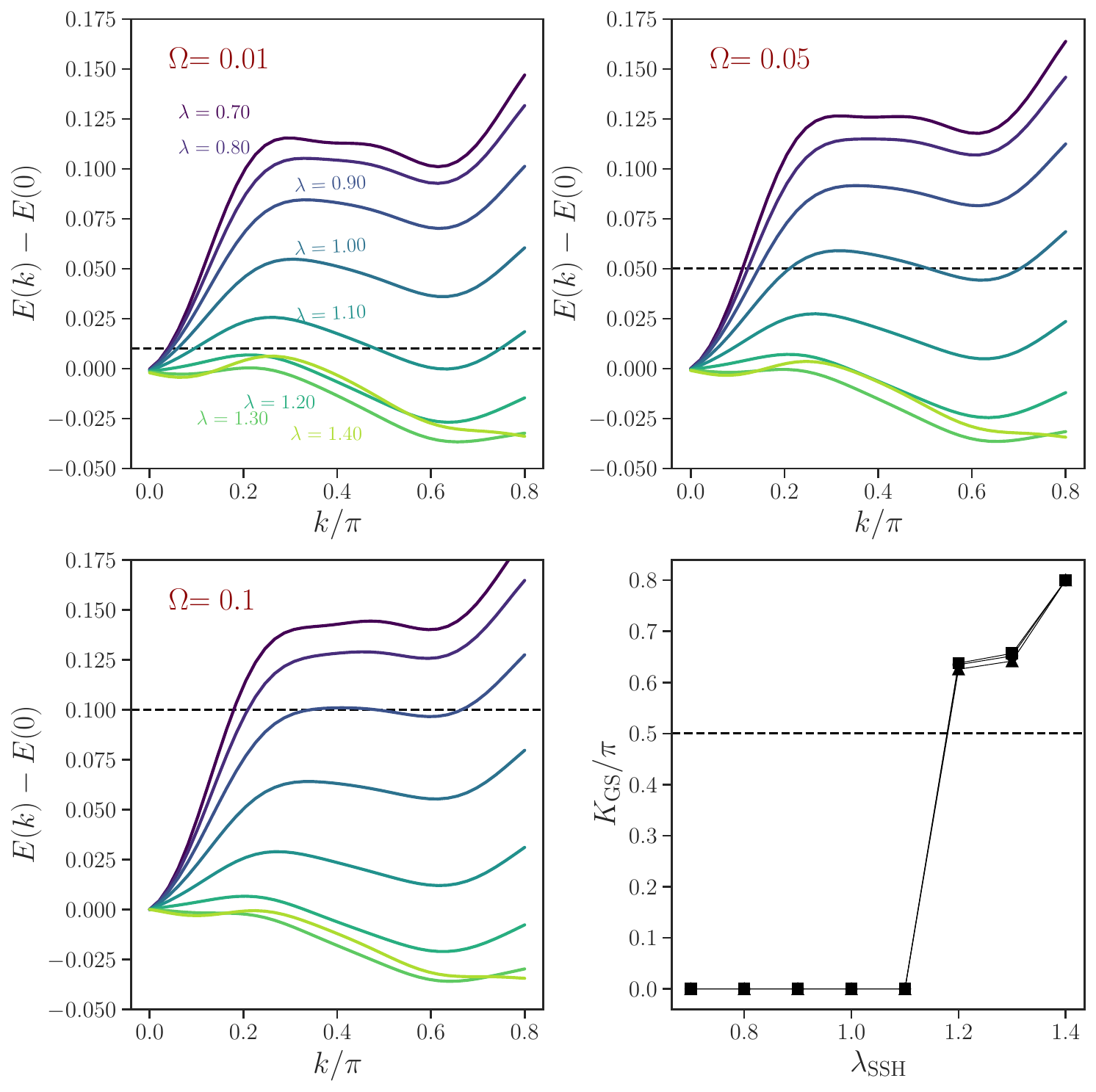}
  \caption{Extrapolation of the polaron dispersion relation $E_{\rm P}$ in phonon frequency space ($\Omega$) for the pure SSH model. The dispersion curves are predicted by ML models trained at phonon frequencies $\Omega \in [0.4,0.9]$. The solid lines are the predictions of the model for various coupling strengths. The horizontal dashed lines show the limit of the dispersion bandwidth equal to $\Omega$. Lower right: The dependence of $K_{\rm GS}$ of the SSH polaron on $\lambda_{\rm SSH}$ at three phonon frequencies corresponding to the dispersion curves shown in the other panels.
 }
  \label{extreme_adiabtic-2}
\end{figure*}

Given that ML predictions are accurate for $\Omega=0.4$, we now 
test the limits of ML models in predicting polaron dispersions in the extreme adiabatic regime. 
Previous work showed that $K_{\rm GS}$ undergoes the transition to  
$\pi/2$ for all frequencies above $0.4$ \cite{DominicMona}. However, the transition at the lowest considered phonon frequencies  is predicted to
occur at very large coupling strength $\lambda_{\rm SSH} \sim 100$ \cite{DominicMona}. 
Our results of Figure \ref{ssh_low_04} show that the transition may happen at much lower values of $\lambda_{\rm SSH}$. 

We include the GGCE results into the training set for another set of ML models. In the process of training these ML models, we have found that including data from a narrower range of frequencies, just above the prediction range, leads to better quantitative predictions. 
This happens because the resulting models are less biased by the different physics at high phonon frequencies. Therefore, we produce a new set of ML models by training with the polaron dispersions in the phonon frequency range
$\Omega \in [0.4,0.7]$. We predict the polaron dispersions at $\Omega=[0.4,0.3,0.2,0.1,0.01]$ as shown in Figs. \ref{extreme_adiabtic-1} and  \ref{extreme_adiabtic-2}.

These predictions are in the phonon frequency regime that is currently out of reach of rigorous quantum calculations. They can therefore not be independently tested. 
However, there are some general physical features that one can verify to confirm the validity of the ML predictions. We observe that the dispersion curves predcited by ML 
for $\Omega=0.4$ and $\Omega = 0.3$ satisfy $|E_P(k)-E_P(0)| \leq \Omega$. This condition is also approximately satisfied for $\Omega=0.2$. We  further observe that the dispersion curves predicted 
for $\Omega<0.2$ do not follow this expected behaviour. 

Another way to confirm the validity of ML predictions is by training multiple models with different distributions of training data (e.g. data from different regions of the $[\lambda, \Omega] $ diagram). 
All predictions in Figs. \ref{extreme_adiabtic-1} and  \ref{extreme_adiabtic-2} are qualitatively robust to variations of the training data. 
We thus attach a high level of confidence to the ML predictions for $\Omega \geq 0.3$. The dispersion curves predicted by ML at $\Omega \geq 0.3$ exhibit the sharp $K_{\rm GS} = 0 \leftrightarrow K_{\rm GS} >0$ transition. 
The ML predictions at $\Omega < 0.3$ cannot be regarded as reliable. 
Interestingly, our results indicate a qualitative change in the ML predictions at the phonon frequency around 0.3. The breakdown of ML predictions below $\Omega = 0.3$ may be indicative of a qualitative change in polaron physics. This is consistent with the DMRG results of 
Grundner  et al  at $\Omega = 0.2$. \cite{grundner2023cooper}.

\begin{figure*}
   {\includegraphics[width=0.45\textwidth]{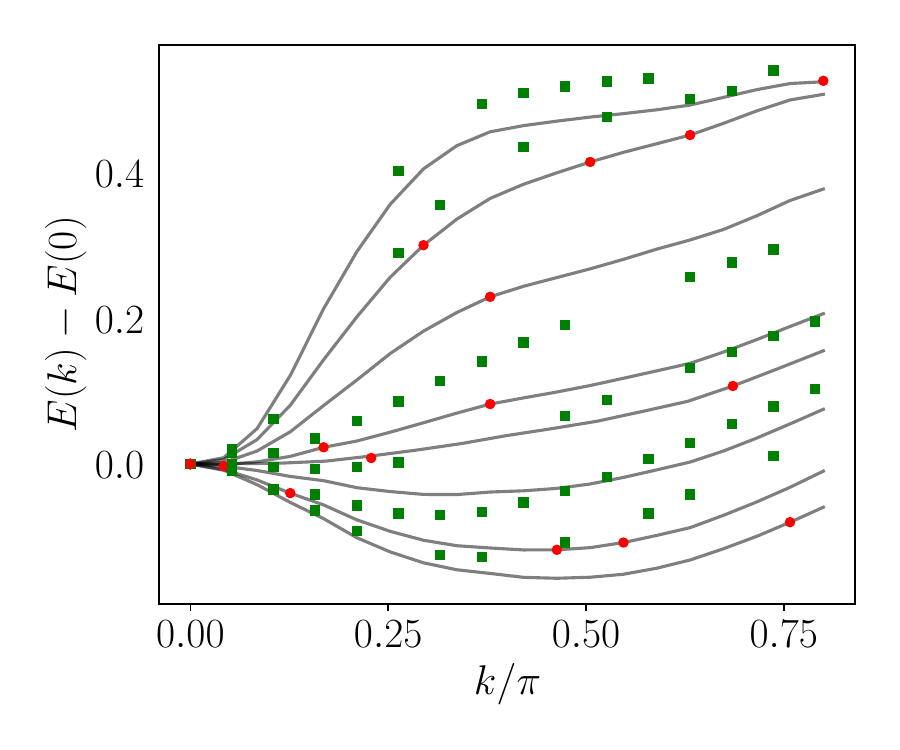}}
   {\raisebox{4mm}{\includegraphics[width=0.45\textwidth]{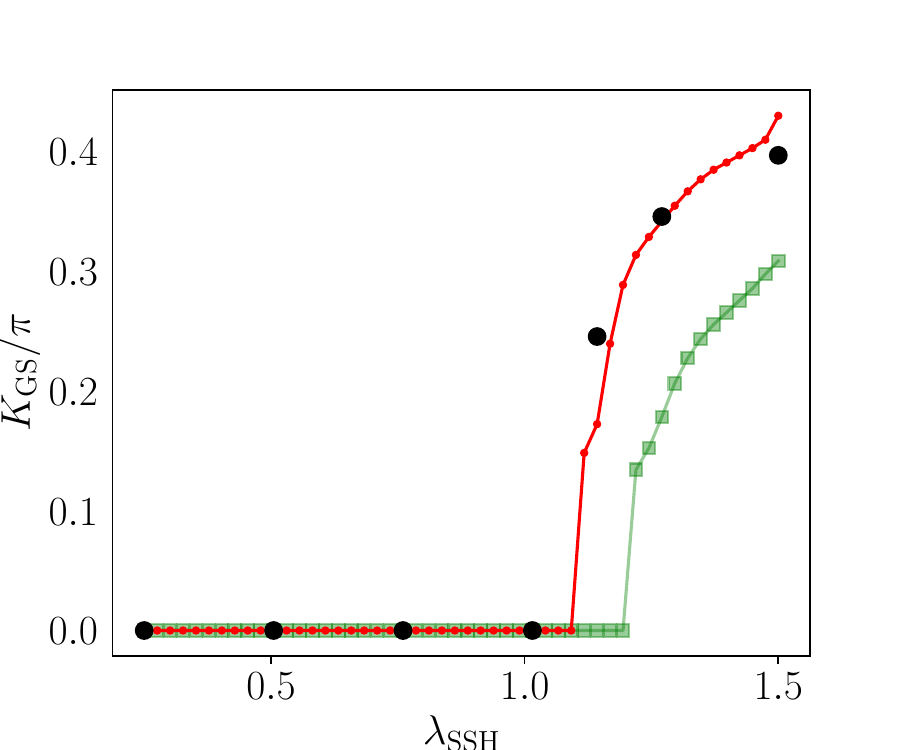}}}
    {\includegraphics[width=0.9\textwidth]{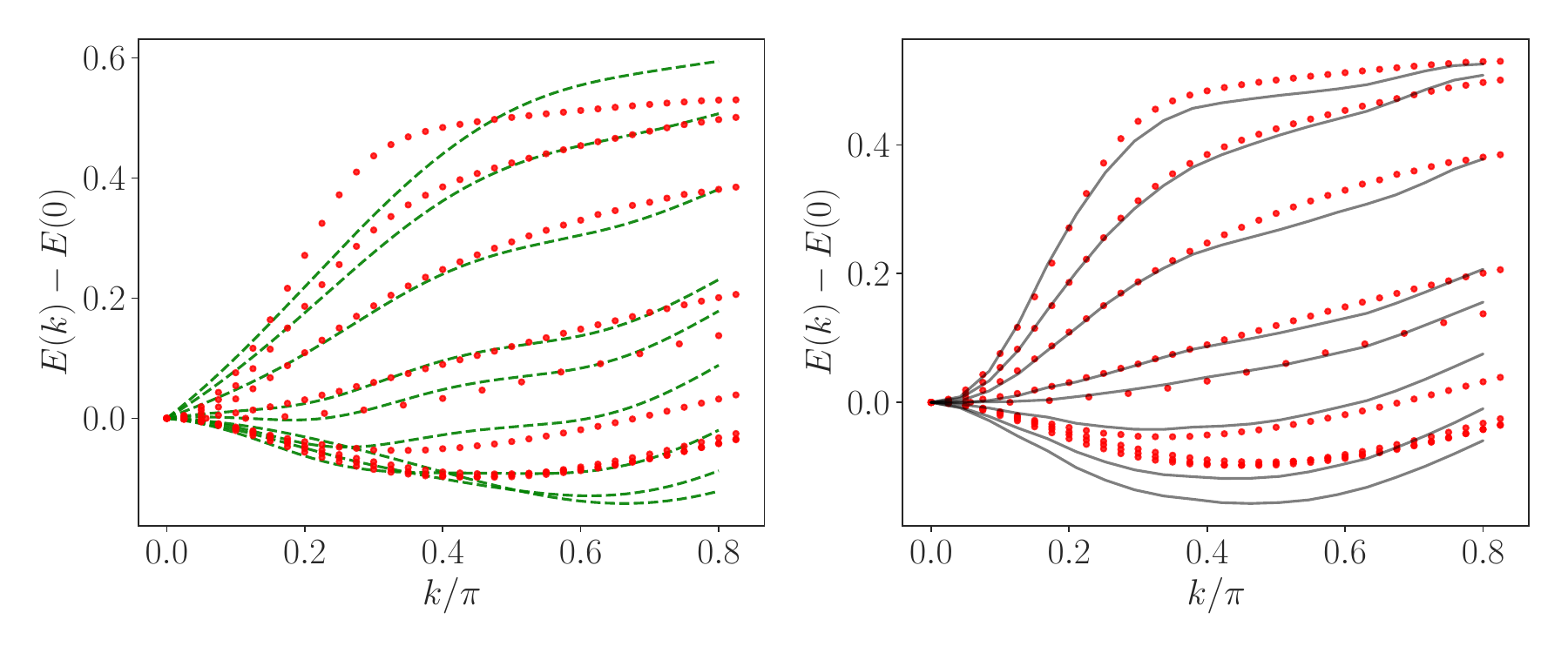}}
    \caption{Left Top:  The green squares (red circles) show the total of 100 (15)  points sampled from the lower (higher) fidelity quantum calculations and the curves are the NARGP model predictions trained on this combination of points. The circles represent expensive quantum calculations for $\Omega=0.5$. 
    The solid lines are evaluated for $\lambda_{\rm SSH}\in [0.25,1.75]$ and $k\in [0,0.8\pi]$. 
    Right Top: Ground-state momentum $K_{\rm GS}$ of the SSH polaron as a function of $\lambda_{\rm SSH}$. The red circles and the green squares are the NARGP model predictions for high fidelity data with the sharp transition at $\lambda^{c}_{\rm SSH}=1.117$ and low fidelity data with the transition at $\lambda^{c}_{\rm SSH}=1.121$, respectively. The black symbols are the GGCE calculations presented for reference.
    Left Bottom: Red circles -- rigorous calculations (reference); broken curves -- predictions of ML models trained directly by 15 calculations
    Right Bottom: Red circles -- rigorous calculations (same as on the left); solid curves -- NARGP model predictions.
    The difference between the dashed green curves and the solid red curves represents the improvement of accuracy due to combination of low-fidelity and high-fidelity data.  }
  \label{figmultifid_1}
\end{figure*}

\subsection{\label{sec-results_d} Interpolation in variational space}

In this final section, we demonstrate multi-fidelity models that can be used to enhance the efficiency and accuracy of 
the GGCE calculations described in Section \ref{sec-model_ggce}. As mentioned previously, we generalize the inputs to our machine learning models to include $N$ and $M$ of the GGCE calculations, and the outputs to describe the results of quantum calculations at different levels of theory. 
One can view the resulting ML models as surrogate models of the polaron energy in the space of $\left [ \Omega,\lambda_{\rm SSH},k\right]$ as well as the parameters of the GGCE calculations $[N,M]$.

We chose Eq. (\ref{base2}) as our kernel function in the NARGP scheme. 
NARGP models are first trained on 100 points sampled across the $[\lambda_{\rm SSH},k]$ grid, with the polaron dispersions calculated using $(M, N)=(2,4)$, far smaller values than required for convergence. With this set of parameters, the full diagram of the polaron energies in the $[\lambda_{\rm SSH},k]$ can be computed on an 8-core iMac with 16 GB RAM in less than ten minutes. The poor accuracy of this fidelity level can be seen in the green sharp transition curve in Fig. \ref{figmultifid_1}, as the critical coupling strength $\lambda^{c}_{\rm SSH}$ at this level is shifted to 1.21 from the exact value $\lambda^{c}_{\rm SSH}=1.117$. The poor accuracy of this calculation is also evident from the departure of green symbols (circles) from the fully converged calculations represented by red circles in Fig. \ref{figmultifid_1}. However, GP models of these approximate calculations can serve as a very good prior into models of the more accurate results.

We show this by subsequently sampling 15 points from the high-fidelity model with the polaron energies calculated using $(M, N)=(3,9)$, which correspond to fully converged calculations. With this choice of parameters, full GGCE calculations of the entire diagram
 would require approximately 10 hours to generate the illustration shown in Fig. \ref{figmultifid_1} on the same computing device. 
 The difficulty of this calculation quickly scales up with $M\rightarrow M+1, N\rightarrow N+1$ and becomes intractable at $\sim (10,8)$ for most supercomputers  \cite{ggce_main, ggce_code}. 
The results of Figure \ref{figmultifid_1} show that these calculations can be combined to produce predictions for the polaron energies at the high-level theory level with a small fraction of the computation cost. 
The improvement of the prediction accuracy by the inclusion of cheap, low-level calculations is seen in the comparison of the lower panels in Figure  \ref{figmultifid_1}. The left panel illustrates the ML predictions from direct interpolation of the small number of exact polaron energies, whereas the right panel illustrates the predictions of multi-fidelity NARGP models. The computation time to produce the dispersion curves in both panels is very similar, as it is determined by the computation of the 15 polaron energies at the  $(M, N)=(3,9)$ level.


\section{\label{sec-conclusion}Conclusion}

The present work demonstrates that it is possible to obtain accurate predictions of polaron properties at low phonon frequencies by  machine learning models trained on polaron properties at high phonon frequencies. This represents an example of machine learning for a quantum problem, where data from an effectively lower-dimensional Hilbert space are used to make predictions for an effectively higher-dimensional Hilbert space. There are numerous examples in quantum theory, where this strategy can be used to extend the range of quantum predictions beyond the limitations of numerical calculations.  
In the present work, we consider the sharp transition in the ground-state momentum of the SSH polaron and examine the evolution of this transition from the anti-adiabatic regime to the adiabatic regime. Whether the sharp transition observed in the anti-adiabatic regime of high phonon frequencies occurs at reasonable electron-phonon coupling in the adiabatic limit has been debated in the recent literature. All of the ML models in the present work predict the sharp transition, even at very low phonon frequencies. 

Extrapolation, by definition, cannot be rigorously tested, until new results become available, whether from theory or experiment. Our conclusions rely on three results. 
First, we test the prediction accuracy of our ML models in the range of phonon frequencies accessible by quantum calculations.  The calculations for this range of phonons 
show that Bayesian ML models proposed here can produce quantitative predictions by extrapolation in the phonon frequency domain. 

For the range of phonon frequencies outside of available rigorous results, we discuss the validity of our ML predictions based on two criteria: criterion (i) --  ML predictions must satisfy basic physical principles;  criterion (ii) -- models trained with different distributions of data 
must produce consistent predictions. The predictions of the sharp transition for the SSH polaron at $\Omega > 0.3$ presented in this work are robust to changes of the training data distributions and should thus be viewed as qualitatively, if not quantitatively,  correct. This is best exemplified by the two panels of Figure 2 that present ML predictions with training data sampled from two very different parts of the Hamiltonian parameter space. 

We observe that ML predictions of polaron dispersions yield energies within the expected band limits for phonon frequencies just beyond the range of the training data. However, as the prediction range of phonon frequencies is reduced to below $\Omega = 0.3$, polaron dispersions predicted by ML become unphysical and exceed the allowed energy range. This signals the onset of the break-down of the ML predictions. This may be the result of the qualitative change in polaron physics, as suggested by the work in Ref.  \cite{grundner2023cooper}. 
While the present article was in preparation, the authors of Ref. \cite{grundner2023cooper} reported the DMRG calculations of the SSH bipolaron energy for a lattice of 256 sites with phonon frequency $\Omega = 0.2$. The authors find that the sharp transition does not occur at this phonon frequency in the range of considered values of $\lambda_{\rm SSH} \in [0, 1.5]$. An unstable bipolaron is formed, which decays as the electron - phonon coupling strength is increased. This suggests departure from the trends observed in Refs. \cite{DominicMona,PhysRevLett.121.247001,roman_sous_bp,sous2017phonon}, where stable bipolarons exist at strong coupling.


We have also demonstrated Bayesian models that use the posterior distributions of Gaussian processes based on highly approximate quantum calculations as the prior distribution for models of more accurate quantum results. 
This drastically reduces the number of fully converged calculations required to map out the full dispersion relations for the full range of Hamiltonian parameters of interest. For example, we have demonstrated that full dispersion relations for the SSH polarons in the range of electron - phonon couplings from 
$\lambda_{\rm SSH}=0.25$ to $1.75$ producing the correct evolution of the sharp transition in polaron ground-state momentum for $\Omega = 0.5$
can be obtained with only 15 polaron energy calculations. This strategy can be employed to build accurate surrogate models of quantum results for problems with a hierarchy of approximate methods, where the number of required quantum calculations decreases with the numerical complexity of the method. The loss of accuracy is fully controlled and can potentially be reduced to negligible errors by increasing the number of calculations  as well as by aligning the kernels of the underlying models with the most accurate results.

\begin{acknowledgments}
This work was supported by NSERC of Canada. We also acknowledge financial support from the Quantum Electronic
Science \& Technology (QuEST) Award from the Stewart Blusson Quantum Matter Institute.
\end{acknowledgments}

\nocite{*}

\bibliography{apssamp}

\end{document}